\documentstyle[amssymb,aps,prl,epsfig,multicol]{revtex}

\begin{document}
\title{Superconducting properties of well-shaped MgB$_2$ single crystal}
\author{Kijoon H. P. Kim,$^{1}$  Jae-Hyuk Choi,$^{1}$ C. U. Jung,$^{1}$
P. Chowdhury,$^{1}$ Min-Seok Park,$^{1}$ Heon-Jung Kim,$^{1}$ J.
Y. Kim,$^{1}$ Zhonglian Du,$^{1}$ Eun-Mi Choi,$^{1}$ Mun-Seog
Kim,$^{1}$ W. N. Kang,$^{1}$ Sung-Ik Lee,$^{1}$~\cite{silee} Gun
Yong Sung,$^{2}$ and Jeong Yong Lee$^{3}$}

\address{$^{1}$National Creative Research Initiative Center for Superconductivity and\\
Department of Physics, Pohang University of Science and Technology, Pohang\\
790-784, Republic of Korea\\
$^{2}$Telecommunications Basic Research Laboratory, Electronics and\\
Telecommunications Research Institute, Taejon 305-350, Republic of Korea\\
$^{3}$Department of Material Science and Engineering, Korea Advanced\\
Institute of Science and Technology, Taejon 305-701, Republic of Korea}
\date{\today}
\maketitle

\begin{abstract}
We report measurements of the transport and the magnetic
properties of high-quality, sub-millimeter-sized MgB$_{2}$ single
crystals with clear hexagonal-plate shapes. The low-field
magnetization and the magnetic hysteresis curves show the vortex
pinning of these crystals to be very weak. The Debye temperature
of $\Theta_{D}\sim 1100$ K, obtained from the zero-field
resistance curve, suggests that the normal-state transport
properties are dominated by electron-phonon interactions. The
resistivity ratio between 40 K and 300 K was about 5, and the
upper critical field anisotropy ratio was 3 $\pm$ 0.2 at
temperatures around 32 K.
\end{abstract}


\begin{multicols}{2}
The recent discovery~\cite{akimitsu_2001} of binary metallic
MgB$_2$ with a superconducting transition temperature of 39 K has
attracted great
scientific~\cite{kang_hall,chen,karapetrov,lima,zhu,xue,jung_pc,canfield_prl,finnemore_prl,budko,MHJung,kong,jung_apl,xu,lee,budko_prl,walti,kim}
and applicatory
interests~\cite{bugoslavsky1,labalestier,hpkim,kang_science,eom,bugoslavsky2}.
The negligibly small effect of its grain boundary on the
supercurrent~\cite{bugoslavsky1,labalestier,hpkim} suggests
increased potential for device applications. Also, vortex pinning,
and thus the critical current, is vastly enhanced in
$c$-axis-oriented thin films~\cite{kang_science,eom} or in the
bulk when disorder is induced by proton
irradiation~\cite{bugoslavsky2}. Aside from the basic properties
such as the charge carrier type~\cite{kang_hall}, many scientific
issues, such as the order parameter
symmetry~\cite{chen,karapetrov}, the upper critical field
anisotropy ratio~\cite{lima,MHJung},
$\gamma=H_{c2}^{ab}/H_{c2}^{c}$, the $\Theta_D$~\cite{poole}, and
the transport properties of the normal-state are still
controversial.

Especially, the anisotropy is an important property because it
significantly affects the electronic and the magnetic properties,
such as the pinning mechanism of this material. The anisotropy
ratio has been reported to be 6 - 9 for the powder. This range of
values was estimated by using conduction electron spin
resonance~\cite{simon}. The values for aligned
crystallites~\cite{lima}, $c$-axis oriented
films~\cite{MHJung,patnaik}, and single crystals grown by
different techniques~\cite{xu,lee} are reported as 1.7, 1.3 - 2,
and 2.6 - 2.7, respectively.

The high superconducting transition temperature in MgB$_2$ has
been considered to be due to strong electron-phonon
coupling~\cite{budko_prl,Kortus,Monteverde}. Thus it is important
to know whether the normal-state transport properties can be
described by a simple electron-phonon interaction, or electron
correlation has to be taken into account.

Other issues are the temperature dependence of the normal-state
resistivity and the value of the residual resistivity ratio (RRR),
$\rho$(300 K)$/\rho$(40 K). The RRR was reported to vary from 2 to
25 depending on the preparation
conditions~\cite{zhu,xue,jung_pc,canfield_prl,finnemore_prl,kang_science,eom},
and the magnetoresistance (MR) at normal-state was reported to
vary from 1\% to 60\%~\cite{jung_pc,finnemore_prl,budko,MHJung},
with a rough correlation between higher RRR and higher MR values.
These issues can be clarified if these quantities are measured in
very clean single crystals. Such results will also help to
construct a theoretical formulation~\cite{kong}.

In this Letter, we report the transport and the magnetic
properties of very clean MgB$_2$ single crystals. The crystals
were found to have well-shaped hexagonal plates with an $a$-axis
lattice constant of 3.09 \AA. The superconducting transition
occurred at 38 K with a sharp transition width of 0.3 K. The
low-field magnetization and the magnetic hysteresis curve showed
the vortex pinning to be very weak, which supported our crystals
being very clean. The $\gamma$ and the $\Theta_D$ were obtained by
directly measuring the temperature- and field-dependence of
resistance for different field directions.

MgB$_2$ bulk pieces~\cite{jung_pc,jung_apl} were heat treated in a
Mg flux inside a Nb tube, which was sealed in an inert gas
atmosphere by using an arc furnace. Then, the Nb tube was put
inside a quartz ampoule, which was sealed in vacuum. The quartz
tube was heated for one hour at 1050 $^\circ$C, then very slowly
cooled to 700 $^\circ$C for five to fifteen days, and quenched to
room temperature. For all measurements, the single crystals were
separated from the resultant matrix using a mechanical method.
Details of growth will be found elsewhere~\cite{jung_choi}. The
crystal images were observed using a polarizing optical microscope
and a field-emission scanning electron microscope (SEM).
Structural analysis was carried out using a high-resolution
transmission electron microscope (HRTEM). The magnetization curves
were measured by using a SQUID magnetometer. Resistivity
measurements were performed using the standard dc 4-probe method.

\begin{figure}
\centering \epsfig{file=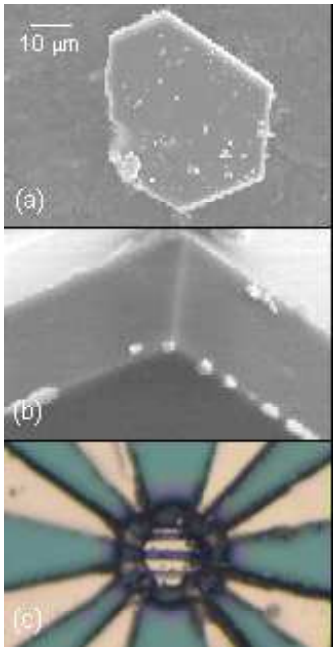, width=6cm} \vskip 0.5 cm
\caption{(a) SEM image of a hexagonal, thin plate with a size of
about 50 $ \protect\mu $m, (b) A magnified view of the upper
corner of the crystal shows smooth surfaces and sharp edges. The
white spots at the edge were stuck weakly on the crystal surface
and were about 100 nm in diameter. (c) Optical microscope image of
the 4-probe contact leads which were made on a single crystal by
using a photolithography technique.} \label{fig1}
\end{figure}

Figure~\ref{fig1}(a) shows a typical SEM image for a MgB$_{2}$
single crystal. Most of the crystals were found to have
hexagonal-plate forms with typical edge angles of 120 degrees and
very flat surfaces, which were very shiny while observed using a
polarizing optical microscope. The crystals were about $20\sim 60$
$\mu$m in diagonal length and $2\sim 6$ $\mu$m in thickness. A
recent study~\cite{zhu} showed that [001] twist grain boundaries,
formed by rotations along the $c$-axis (typically by about 4
degrees), were the major grain boundaries in polycrystalline
MgB$_{2}$, which could be attributed to the weaker Mg-B
bonding~\cite{zhu}. HRTEM and SEM studies showed no grain
boundaries in our crystals. Figure~\ref{fig1}(b) shows a magnified
view of the upper corner of the crystal shown in
Fig.~\ref{fig1}(a). The smooth surfaces and the sharp edges in
this figure confirm that our small crystals had the least
probability of having mosaic aggregates of nanocrystals either
along the $ab$-plane or along the $c$-axis; thus, we had a better
chance to study their intrinsic properties with the help of a
micro-fabrication technique.

To measure the temperature and the field dependencies of the
in-plane resistivity of the crystal, we fabricated four electrical
metal leads (bright area) on the top surface of the crystal, as
shown in Fig.~\ref{fig1}(c). For lithography, selected MgB$_2$
single crystals were fixed on oxidized Si substrate by holding
their hexagonal edges using a negative photoresist (OMR-83, TOK
Co. Ltd.) as glue. They were then soft-baked. After the surfaces
of the samples were cleaned by using Ar-ion milling with a beam
voltage of 350 V and a beam current of ~0.2 mA/cm$^2$, four
electrical pads were patterned using a positive photoresist (AZ
7210, Clariant Industries Ltd.). Three metal layers, a 100
nm-thick Ti film, a 1000 nm-thick Ag film, and a 100 nm-thick Au
film,  were deposited in sequence after another ion milling
treatment. The contact resistances were less than 2 $\Omega$, and
the approximate distance between the voltage pads was about 7
$\mu$m. The bias current for the resistance measurement was 0.1
$\sim$ 0.2 mA, which, judging from the current-voltage
characteristics, was in the ohmic range.

\begin{figure}
\centering \epsfig{file=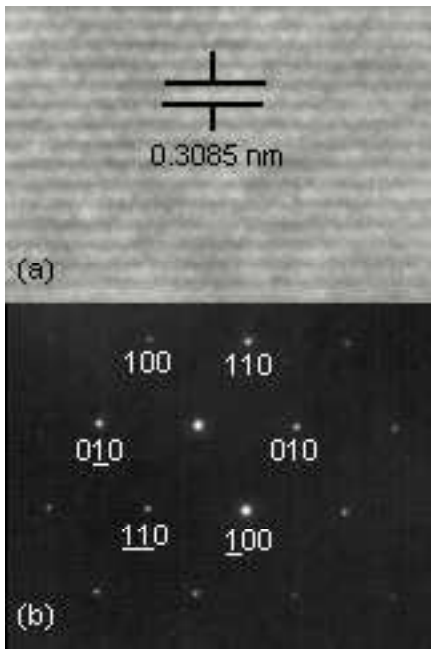, width=6cm} \vskip 0.5 cm
\caption{(a) HRTEM image of a MgB$_2$ single crystal and (b)
selected area electron diffraction pattern for a beam direction of
[001] in the hexagonal structure.} \label{fig2}
\end{figure}

To confirm the structure of the MgB$_{2}$ phase, we took a
plane-view HRTEM image, as shown in Fig.~\ref{fig2}. From this
high-resolution image, the $a$-axis lattice parameter was found to
be $3.09 \pm 0.06$ \AA, which was consistent with the value
determined from X-ray powder diffractometry performed on
polycrystalline samples~\cite{akimitsu_2001}. Figure~\ref{fig2}(b)
shows the electron diffraction pattern in a selected area for a
beam direction of [001]. This result clearly indicates that this
crystal has the hexagonal structure of MgB$_{2}$.

\begin{figure}
\centering \epsfig{file=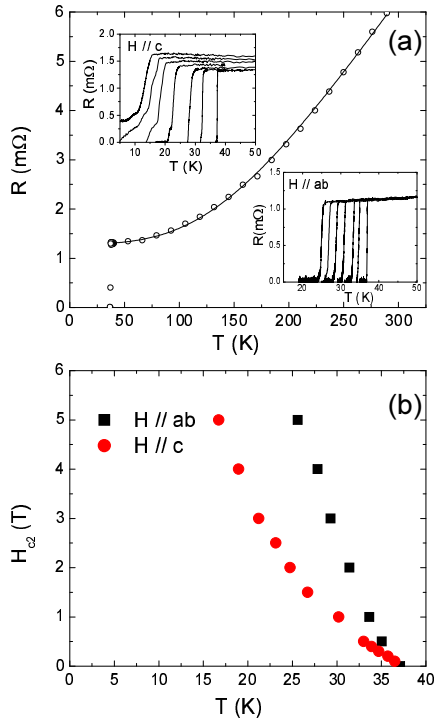, width=6cm} \vskip 0.5 cm
\caption{(a) Resistance of a MgB$_2$ single crystal as a function
of temperature for magnetic fields from 0 to 5 T. The residual
resistivity ratio was about 5. The insets show the resistance for
fields perpendicular and parallel to the crystal $c$-axis at 0.0,
0.5, 1.0, 2.0, 3.0, 4.0, and 5 T. (b) The upper critical field
determined from a 10\% drop of the resistance as a function of
temperature for several fields up to $H = 5$ T.} \label{fig3}
\end{figure}

Figure~\ref{fig3}(a) shows the in-plane resistance as a function
of temperature. A superconducting transition appeared near 38 K,
and the transition width was 0.3 K based on the 10 to 90\% drop of
the resistance curve at zero field. The RRR was about 5, which was
confirmed for our several crystals and was consistent with the
values for single crystals recently reported~\cite{xu,lee}. This
is quite different from the reported values for
polycrystals~\cite{zhu,xue,jung_pc,canfield_prl,finnemore_prl} and
high-quality thin films~\cite{kang_science,eom}. Extrinsic
effects, such as impurities or grain boundaries, might be the
origins of these diverse observations~\cite{zhu,xue}. The {\it
c}-axis resistivity of single crystals should be studied for more
understanding; however, this is still challenging yet due to the
thickness of our crystals.

The solid line in Fig.~\ref{fig3} is a fitting curve obtained
using the Bloch-Gr\"{u}neisen formula~\cite{poole} in the normal
state: $R(T)=R_0+R_{\rm ph}(T)$ where $R_0$ is the
temperature-independent residual part and $R_{\rm ph}(T)$ the
phonon scattering contribution given by the relation:
\begin{equation}
R_{\rm ph}(T)= R_1
\left(\frac{T}{\Theta_D}\right)^m\int_0^{\Theta_D/T}\frac{z^mdz}{(1-e^{-z})(e^z-1)},
\end{equation}
with $R_1$ being a proportionality constant. The best fit to our
data was obtained with $m$ = 3 and $\Theta_D \sim 1100$ K. This
values of the $\Theta_D$ is comparable to those (746 - 1050 K)
previously reported based on the specific heat and resistivity
measurements on polycrystalline
samples~\cite{budko_prl,walti,Kremer,Bouquet,Putti}. This result
suggests that the normal-state transport properties are well
described by an electron-phonon interaction without taking an
electron-electron interaction into account.

The two insets of Fig.~\ref{fig3}(a) show the field dependencies
of the in-plane resistances while maintaining the applied field
perpendicular to the current path. The magnetoresistance at 5 T
was found to change with the direction of the field with respect
to the crystal axis. For fields parallel to the $ab$-plane
($H\parallel ab$) and to the $c$-axis ($H\parallel c$), the
magnetoresistances at 40 K were $\lesssim 3$\% and $\sim 20$\%,
respectively, which was confirmed for our several crystals. To
determine the temperature dependence of the upper critical field
$H_{c2}$, we measured the resistance at several fields up to 5 T.
From these curves, we obtained $ H_{c2}(T)$ as shown in
Fig.~\ref{fig3}(b), where a 10\% drop of the resistance was
adopted to determine $T_{c}(H)$. The ratio of the upper critical
field for $H\parallel ab$ to that for $H\parallel c$ was 3 $\pm$
0.2 at temperatures around 32 K. This value is consistent with
those for single crystals grown by different
techniques~\cite{xu,lee}.

To confirm the bulk properties of the superconductivity, we
measured the low-field magnetization curve $M(T)$ and the magnetic
hysteresis curve $M(H)$. Since the volume of one crystal is rather
small, we fixed about a hundred single crystals on a Si substrate
with their $c$-axes aligned perpendicular to the substrate
surface, which is similar to the method used by
others~\cite{lima}. To avoid spurious signals from matrix, we used
an optical microscope and a sample-handing device equipped with a
precision $xyz$-stage and a micro-tip to collect large single
crystals one by one.

\begin{figure}
\centering \epsfig{file=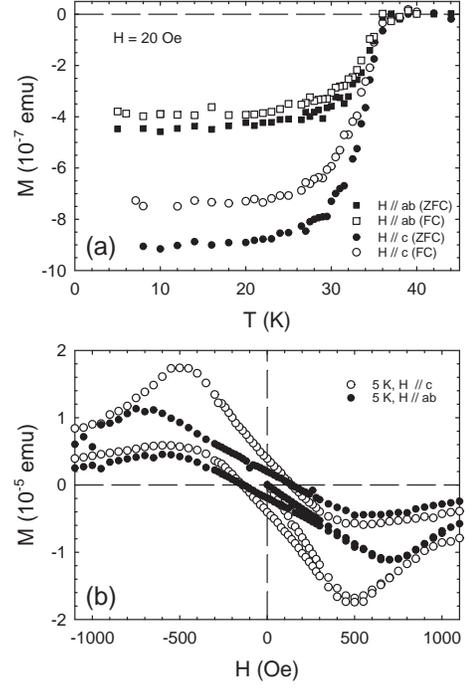, width=6cm} \vskip 0.5 cm
\caption{(a) Low-field $M(T)$ curves of MgB$_2$ single crystals
measured at 20 Oe for fields parallel and perpendicular to the
$c$-axis. (b) $M(H)$ hysteresis curves measured at 5 K. }
\label{fig4}
\end{figure}

Figure~\ref{fig4}(a) shows the $M(T)$ curves measured at 20 Oe in
the zero-field-cooling (ZFC) and the field-cooling (FC) modes. The
$T_c$ onset was observed to be $\sim 38$ K, which is consistent
with the value obtained from the resistance measurement. The
difference between the FC and the ZFC data is quite small compared
to those of polycrystalline samples~\cite{jung_pc,jung_apl} or
single crystals prepared at higher temperature~\cite{lee},
suggesting that pinning is very weak in our single crystals. The
different values of $M(T)$ for different field directions give a
demagnetization factor $D\gtrsim0.6$, which is consistent with the
value calculated by considering the shape of the crystals.

Figure~\ref{fig4}(b) shows the magnetic hysteresis curves $M(H)$
at 5 K. The $ M(H)$ data show a negligible paramagnetic
background. The different slopes of the $M(H)$ curves at low
starting fields ($H < 300$ Oe) for different field directions are
due to the demagnetization factors being different. The difference
between the values of $M(H)$ for the increasing and the decreasing
field branches is rather small, which also shows that bulk pinning
is very small and, thus, supports our crystals being very clean.
The results in Figs.~\ref{fig4}(a) and~\ref{fig4}(b) indicate that
the strong bulk pinning reported for polycrystals~\cite{kim} and
thin films~\cite{kang_science,eom} is due to extrinsic pinning
sites, such as grain boundaries and crystallographic defects.

In summary, we report the transport and the magnetic properties
for  high-quality MgB$_2$ single crystals. A superconducting
transition occurred at 38 K with a sharp transition width of 0.3
K. The low-field magnetization and the magnetic hysteresis curve
showed the vortex pinning to be very weak. From the resistance
measurement, a $\Theta_D$ of $\sim 1100$ K was obtained using the
Bloch-Gr\"{u}neisen formula, which suggests that the normal-state
transport properties are dominated by an electron-phonon
interaction rather than by an electron-electron interaction. A RRR
of 5 and a $\gamma$ of 3 $\pm$ 0.2 at temperatures around 32 K,
were obtained.

This work is supported by the Ministry of Science and Technology
of Korea through the Creative Research Initiative Program. This
work was partially supported by the National Research Laboratory
Program through the Korea Institute of Science and Technology
Evaluation and Planning. We acknowledge Do Hyun Lim and Taek-Jung
Shin at Iljin Diamond Co., Ltd., for their help.


\newpage

\end{multicols}

\end{document}